\documentclass[aps,prd,twocolumn]{revtex4-2}
\usepackage{graphicx,epsfig}
\usepackage{amsmath,amssymb}
\usepackage{lineno,hyperref}
\hypersetup{colorlinks=true}
\usepackage{xcolor}

\usepackage{caption}
\usepackage{comment}
\usepackage{relsize}

\renewcommand{\phi}{\varphi}

\begin{document}
\title{Antiglitch: a Quasi-physical Model for Removing Short Glitches from LIGO and Virgo Data}
\author {Ruxandra Bondarescu}
\email{ruxandra@icc.ub.edu}
\affiliation{Institut de Ci\`encies del Cosmos (ICCUB),
Facultat de F\'{i}sica,
 Universitat de Barcelona, Mart\'i i Franqu\`es 1, E-08028 Barcelona, Spain}
 \author{
Andrew Lundgren}
\email{andrew.lundgren@port.ac.uk}
\affiliation{Institute of Cosmology and Gravitation, University of Portsmouth, Dennis
Sciama Building, Burnaby Road, Portsmouth, PO1 3FX, United Kingdom}
\author{Ronaldas Macas}
\email{Ronaldas.Macas@port.ac.uk}
\affiliation{Institute of Cosmology and Gravitation, University of Portsmouth, Dennis
Sciama Building, Burnaby Road, Portsmouth, PO1 3FX, United Kingdom}

\begin{abstract}
Gravitational-wave observatories become more sensitive with each observing run, increasing the number of detected gravitational-wave signals. A limiting factor in identifying these signals is the presence of transient non-Gaussian noise, which generates glitches that can mimic gravitational wave signals.
Our work provides a quasi-physical model waveform for the four most common types of short transient glitches, which are particularly problematic in the search for high-mass black hole binaries. Our model has only a few, physically interpretable parameters: central frequency, bandwidth, phase,  amplitude and time.
We demonstrate the accuracy of this model by fitting and removing a large sample of glitches from a month of LIGO and Virgo data from the O3 observing run. We can effectively remove three of the four types of short transients. We finally map the ability of these glitches to mimic binary black hole signals.
\end{abstract}
\maketitle

\section{Introduction}  
The LIGO, Virgo, and KAGRA Collaborations (LVK) \citep{Harry:2010zz, LIGOScientific:2014pky,acernese2014advanced,aso2013interferometer} have recently started their fourth observing run (O4), which is expected to achieve the highest sensitivity to gravitational wave signals to date. Gravitational wave observatories are designed for events in the tens of Hz to kHz band with minimum noise at $100-200$ Hz. Their sensitivity is described via the power spectral density (PSD) of the stationary noise contributions to the measured gravitational-wave strain, which neglects non-Gaussian and non-stationary contributions.

Glitches are the general name given to disturbances that occur in the strain time series in addition to the stationary Gaussian background \citep{merritt2021transient}. There are a wide variety of types and causes of glitches. We will focus here only on short transient glitches, less than $0.1$ seconds long. These are the most problematic kind, as their origin is generally unknown and they are similar to high-mass binary black hole signals. Thus they degrade the performance of the searches because they can be mistaken for real signals, obscure a real event or impact the sky localization if they occur nearby in time \citep{glitch_pe_1, glitch_pe_2, glitch_pe_3, gw170817_bw, skyloc}.

There are several known classes of glitches which are common in the detectors, and most glitches that occur can be identified as one of these classes. The identification is made by human inspection of the appearance in a time-frequency spectrogram of the strain data. GravitySpy \cite{zevin2017gspy} used citizen science volunteers to train a machine learning model which automatically classifies glitches. We use the database of these classifications as a starting point for our model, and as a convenient way to identify all of the occurrences of a specific class. GravitySpy identifies four classes of short transients, which have been named blips \citep{cabero2019blip}, low-frequency blips, tomtes, and koi fish \citep{bahaadini2018machine}.

These glitches have plagued O3 \citep{davis2022subtracting} and are expected to be more of a problem with the increased sensitivity in O4. They are particularly a problem for searches for binary black holes with high masses ($> 100 M_\odot$) because these are short with few gravitational-wave cycles to provide discriminating power.
The physical origins of these short transients are generally unknown. Hence we do not know how similar glitches in a given class actually are to each other. In the best case, they would all be identical up to an overall amplitude, but we will show that this is not the case. The ability to classify them does however suggest that their waveform can be described by a small number of latent variables. 

In this paper, we build and demonstrate a precise model of these glitches which can be used to identify them with high confidence and also to subtract them from the data. Our model is quasi-physical, meaning that it encodes the characteristic properties of the glitches in a few physically meaningful parameters, but is not derived from a specific physical model of the glitch mechanism. As long as the detector is near its operating point and has not reached analog or digital saturation, the glitch waveform is expected to be additive. Hence, if it can be modelled, it can be subtracted. With the detectors now online, it is urgent to develop methods to remove glitches in real time. 

Our model is data driven and is tested on a month of LIGO and Virgo data from the O3 observing run. It assumes the glitches (1) contain a single peak frequency, (2) are either symmetric or antisymmetric in time, and (3) have a spectrum that is normal in log-frequency.  We can ensure the quick and accurate removal of the blip, low-frequency blip and tomte glitches. Our model has also been tested for koi fish glitches, but they have more complicated morphology making their removal less effective.

Glitches have been investigated in numerous other works \citep{PhysRevD.102.023011, george2018classification,davis2022subtracting,cabero2019blip,tolley2023archenemy}. Many authors propose glitch removal via unmodelled subtraction techniques \citep{cornish2015bayeswave, cornish2021bayeswave,george2018classification}, which do not make specific assumptions about glitch morphology, or make use of neural networks with no physical inductive bias. While these methods are more general and can remove a variety of glitches at the post-processing stage, they are also computationally expensive and cannot be performed fast enough to avoid fake triggers. Neural networks with no physical inductive bias remove power reliably, but can also remove signal or other forms of noise. They use a wavelet basis that requires multiple tiles, each with its own noise, which brings in additional degrees of freedom. 

We also compute the match of these glitches to a bank of compact binary coalescence (CBC) templates, using the IMRPhenomXAS waveform \cite{imrphenom}. We predict based on the glitch parameters of our model which CBC templates will be affected and map this to the different types of observed glitches. This information can be used optimally to discriminate between these types of glitches and actual CBC signals in detector noise.

\section{Causes of Glitches}
Glitches are caused by a wide variety of mechanisms of which some are known. However, for several common-occurring glitches, the cause is still unknown. The possible causes are split into four categories, following the path of the calibration loop \cite{abbott2017calibration}: 
\begin{itemize}
\item Physical Glitches. A sudden motion of one of the test masses will be perceived as a glitch. This can be caused by external seismic motion or vibration and should be detected by Physical and Environmental Monitoring (PEM) sensors. Some examples are the shutter of a camera temporarily mounted near the test mass and ravens pecking at the ice on the Liquid Nitrogen supply pipe. It is also possible for internal mechanisms to cause glitches such as the thermal popping of the mirrors due to the release of stress or stray electric fields and currents. 
\item Optical Sensing Glitches: The relative position of the end mirrors is the primary observable of the detectors -- it measures the gravitational-wave strain \cite{abbott2016gw150914}. This is read out by the intensity of light on the photodiodes at the output port of the interferometer. Anything interfering with the beam of light, e.g. some dust or a spot of damage on the photodiode, can be misinterpreted as a length change. 
\item DARM loop Glitches. The mirrors must be kept near their operating point despite external seismic disturbances. This requires a number of control loops that act on the length and angular degrees of freedom. Here, we concentrate on the DARM loop, which converts the optical sensing above into feedback to the end test masses. This is a digital control loop which shapes the feedback to avoid exciting resonances in the suspensions of the mirrors. Timing slips between the front-end computers have caused DARM glitches  in the past \cite{cabero2019blip}. With faster CPUs and more monitoring, this is unlikely to be a problem in recent observing runs.
\item  Actuation Glitches: The mirror positions and angles are controlled by solenoid coils actuating magnetically on masses above the test masses in suspension. The most sensitive, high-frequency feedback is applied by electrostatic drivers, which use large voltages to apply force to the test mass. Problems in either of these electronic systems could manifest as glitches. Some of the short transients investigated here appear to have an electrical origin. However, they have to be investigated further to understand the exact origin.
\end{itemize}

There are many other mirrors in the detector and many auxiliary lengths and angles which are also controlled. Any glitch in these can affect the optical sensing causing apparent motion in the strain channel. A known glitch of this type is RF45 glitches, from the 45 MHz radio frequency modulation system, which affects many auxiliary degrees of freedom. Typically, these kinds of glitches will be sensed strongly by the auxiliary channel monitors and will therefore have a known origin. Hence unknown glitches are more likely to arise in one of the four systems listed above.

\section{Glitch Detection and Classification}

\subsection{Q Transform}

The tool most often used to detect and inspect glitches in LIGO and Virgo data is the Q transform \cite{chatterji2005thesis}. The Q transform defines a set of minimum-uncertainty tiles which cover the time-frequency plane. These tiles are bisquare windows in the frequency domain, effectively bandpasses defined by their central frequency and bandwidth. The Q transform whitens the data and then decomposes the whitened signal onto the tiles, yielding the excess power in each tile. The basis of tiles is non-adaptive except for allowing different choices of the quality factor $Q$; larger values of $Q$ have less span in frequency (lower bandwidth) and hence more span in time. The result is displayed in a spectrogram of excess power which is called either a Q scan or Omega scan. Only one value of $Q$ is used for the spectrogram, which can be determined either automatically or by the analyst. Low $Q$ values will better resolve short transients at the cost of frequency resolution. While this method has proved to be an effective visualization method for human analysts, it has drawbacks. Like any spectrogram, it discards the phase information and only keeps the amplitudes of each tile, and the basis is not orthogonal so it cannot be used to reconstruct a detailed model of a glitch.

Omicron \cite{robinet2020omicron} is a method of searching for general glitches using the Q transform. It runs on a single detector channel (in our case the calibrated strain) and decomposes the data onto tiles for multiple values of $Q$. Tiles above a given threshold in SNR are retained and clustered over time. Because a glitch will intersect multiple tiles, the results are clustered over time so that a glitch is represented by a single trigger. The trigger is characterized by the SNR and frequency of the tile with the largest SNR in the cluster.

\subsection{Glitch Classes}

We focus on the four most common types of isolated short glitches: blips, low-frequency blips, koi fish, and tomtes. The identification is made by the appearance of their spectrogram in the time-frequency domain. Blip glitches are characterized by a narrow tear-drop shape. They are very short-duration transients  (of the order of 5$-$10 ms) with a broad frequency distribution in the region of maximum sensitivity of the detector. This glitch morphology was first observed in Initial LIGO \citep{PhysRevD.102.023011} and has persisted onto Advanced LIGO and Virgo observation. Tomte glitches are wider, have a triangular shape, and are at a lower frequency. Low-frequency blips have a similar rounded shape to blips but do not reach as high a frequency. Koi fish glitches have a similar bandwidth to blips but have extra noise on either side (including a feature around $60$ Hz which forms the 'fins' of the fish).

\subsection{Gravity Spy}

As an initial glitch classification, we use GravitySpy \citep{zevin2017gspy, bahaadini2018machine}.  The GravitySpy pipeline analyzes any Omicron trigger with an SNR above 7.5; weaker glitches are difficult to identify because they are obscured by noise. The classification is from the appearance of the Q scan. Citizen scientist volunteers have made more than a million classifications and have also proposed new categories of glitches \cite{soni2021discovering}. A machine learning model trained using this information now makes automated classifications, still based on the Q scan.

We take May 2019, during the LVK O3a observing period, as our dataset for this paper. The classifications through O3 are discussed in \cite{glanzer2023data} and the classifications are publicly released on Zenodo \cite{gspy_zenodo}. We consider all GravitySpy classified glitches with a machine-learning confidence score above $50 \%$ (though the confidence is strongly peaked toward $100 \%$.
In the L1 (Livingston) detector, there are $485$ blip, $1544$ low-frequency blip, $2920$ tomte, and $1185$ koi fish glitches.
In H1 (Hanford), there are $738$ blips, $325$ low-frequency blips, $136$ tomte, and $1156$ koi fish.
In V1 (Virgo), there are $264$ blips, $0$ low-frequency blips, $96$ tomte, and $196$ koi fish.
In our further analysis, we will use up to $500$ of each type of glitch from each detector, chosen at random from the sample \citep{glanzer2023data}.

\begin{figure}[!htbp]
\includegraphics[width=1.0\columnwidth]{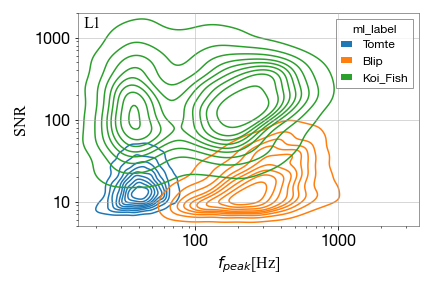}
\includegraphics[width=1.0\columnwidth]{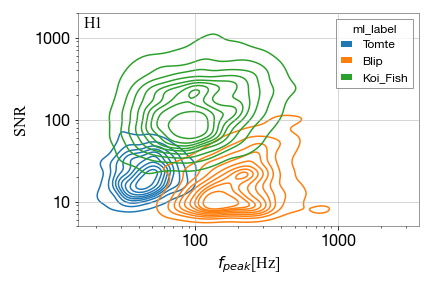}
\includegraphics[width=1.0\columnwidth]{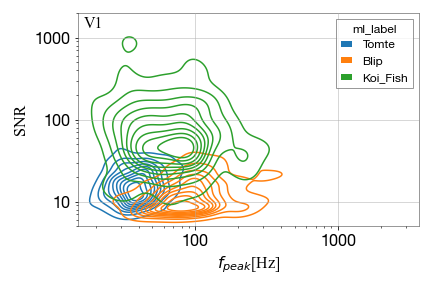}
\caption{The SNR versus frequency distribution for the GravitySpy classifications of tomte, blip and koi fish glitches for L1, H1 and V1.}
\label{fig:GSpyfvsSNR}
\end{figure}

\begin{figure}[!htbp]
\includegraphics[width=1.0\columnwidth]{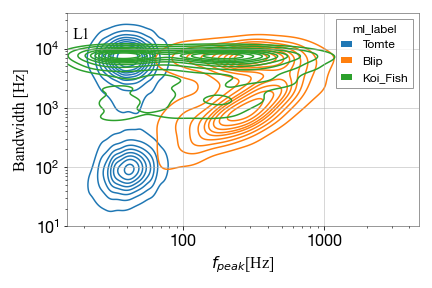}
\includegraphics[width=1.0\columnwidth]{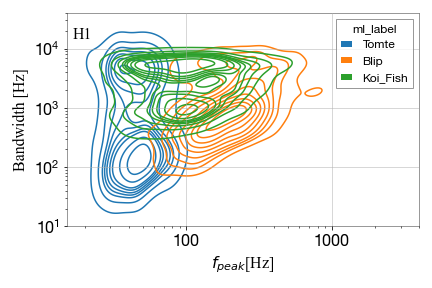 }
\includegraphics[width=1.0\columnwidth]{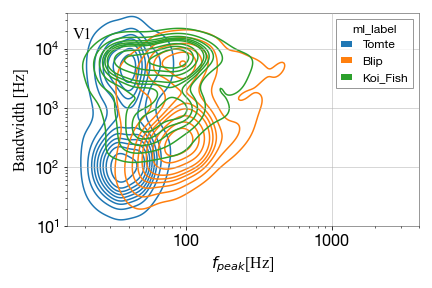 }
\caption{The SNR is plotted as a function of Bandwidth for the Gravityspy classification of tomte, blip and koi fish glitches for L1, H1 and V1. }
\label{fig:GSpyfBandwidth}
\end{figure}

We can begin to see some characteristics of the population. There is a clear excess of tomte and low-frequency blip glitches in L1. Examining the distribution of SNR, frequency, and bandwidth calculated by Omicron demonstrates the different characteristics of the glitch types. As Fig.~\ref{fig:GSpyfvsSNR} shows, koi fish have much higher SNR, while blips and tomtes are similar to each other. The distributions are mostly separated in the frequency-SNR plane. Tomtes have a low peak frequency while blips are higher; koi fish span the range of both. Fig.~\ref{fig:GSpyfBandwidth} shows that only tomte have low bandwidth since they have the highest range of bandwidths. We have suppressed low-frequency blips for clarity, but they are similar to tomtes with lower SNRs.

\section{Glitch Model and Inference}
\begin{figure}[!htbp]
\includegraphics[width=1.0\columnwidth]{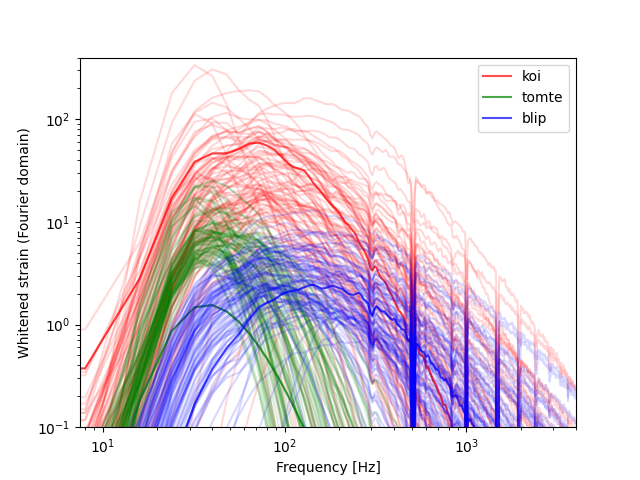}
\includegraphics[width=1.0\columnwidth]{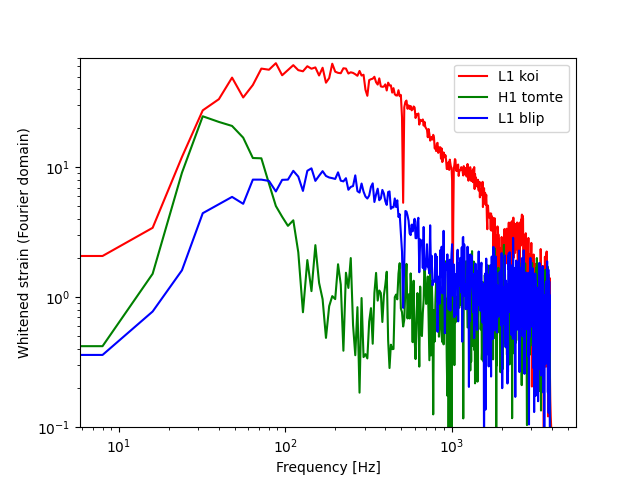}
\caption{The whitened strain of tomte, blip, and koi fish glitches shown in the frequency domain  for data taken in May 2019, during the O3a science run, for a) many glitches in H1 and b) selected glitches in H1 and L1.  It can be seen that they affect the sensitivity of the detector over a wide range of frequencies. At first glance, the koi fish glitch appears to mimic either the blip or the tomte. However, it has a higher amplitude and exhibits some features that cannot always be removed entirely with our simple model.}
\label{fig:FD}
\end{figure}

Fig. \ref{fig:FD}(a) shows the FFTs of the whitened strain for tomte, blip and koi fish glitches (low-frequency blips are omitted for clarity). The whitening process causes dips in the spectra corresponding to strong lines in the instrumental noise. The koi fish and blips both have generally higher central frequencies and wider bandwidths than tomtes. However, the spectra seem to have similar shapes up to scaling and shifting. An example of each glitch type is shown in more detail in Fig. \ref{fig:FD}(b).

\subsection{Quasi-physical Model}

We now propose a simple model for the glitch in the calibrated strain channel. The model is defined in the frequency domain. The whitened spectra are all fairly well matched by a normal in logarithmic frequency, with the parameters varying between different glitches. Because the glitches are so short and have time-symmetric spectrograms, we find that they can be well-represented by a constant phase with respect to frequency. Hence their power will be maximally concentrated around their center if the frequency-domain waveform is constant in phase.  In the frequency domain, the model is
\begin{eqnarray}
    h_0(f) = \mathcal{N}^{-1} \exp\left[-\frac{\gamma}{2} (\log f - \log f_0)^2\right] \\
    h(f) = A \exp\left[i \phi - 2 \pi i f t_0\right] h_0(f) 
\end{eqnarray}
The normalization factor is
\begin{equation}
\mathcal{N} = \sqrt{\mathlarger{\sum}_k \frac{|h_0(f_k)|^2}{S_n(f_k)}}
\end{equation}
with $k$ the index over discrete frequencies so that the template has constant SNR at $A = 1$.

The parameters of the model are the peak frequency $f_0$ and the inverse bandwidth squared $\gamma$. In addition, we need an overall amplitude ($A$), phase ($\phi$), and central time of the glitch ($t_0$). Our data segments will be shifted to put the glitch nearly in the center, but the $t_0$ parameter allows for imperfect centering. A phase of $\phi = 0$ or $\pi$ results in a glitch that is symmetric in time. Conversely, $\phi = \pm \pi/2$ results in a glitch which is antisymmetric in time.

\subsection{Inference of Glitch Parameters}

To determine the parameters of a given glitch, we make use of Bayesian parameter estimation. The log-likelihood for a modelled glitch added to stationary Gaussian noise is
\begin{equation}
\log \mathlarger{\Lambda} = - \mathlarger{\sum}_{k} \frac{|d(f_k) - h(f_k)|^2}{2 S_n(f_k)},
\end{equation}
where $d(f)$ is the data, $h(f)$ the glitch model, and $S_n(f)$ the noise power spectrum, suitably discretized. We use a sample rate of $8192~\mathrm{Hz}$, and our data segment is $1/8~\mathrm{sec}$ around each glitch, with the spectrum estimated by the Welch method on $18$ seconds of data surrounding the glitch, with the central $2$ seconds excised \citep{allen2012findchirp}. The strain data has been scaled by a factor of $10^{23}$ to put it in a reasonable numerical range.

We implement our model in NumPyro \cite{numpyro1, numpyro2} with the model and likelihood implemented in Jax \cite{jax}. We defined a common prior for all glitches in amplitude, time, phase, frequency and bandwidth:
\begin{itemize}
    \item $A \sim \mathcal{N}(0, 400)$
    \item $\phi \sim \textrm{U}(-\pi, \pi)$
    \item $f_0 \sim \mathrm{U}(10, 600)$ [Hz]
    \item $\gamma \sim \mathrm{U}(0.25, 8)$
    \item $t_0 = 0.01 (2 x - 1)$ [sec] ;
        ~ $x \sim \mathlarger{\beta}(2,2)$
\end{itemize}
The text in brackets indicates units where applicable. The time distribution is an affine transform of a Beta distribution chosen to be symmetric around $0$ and of finite extent to prevent wraparound.

We used NUTS sampling \cite{nuts}, which takes less than a minute per glitch on a typical processor. However, the priors are quite broad because they have to encompass a wide range of possible glitches, and we do not yet know the population distribution of these parameters. So we have used a maximum-likelihood estimate for our primary results, which is calculated in three seconds per glitch on a single CPU core.

\section{Results}

\subsection{SNR and Excess Power}

The results of performing inference are a precise model of the glitch under consideration. We can use this to subtract the glitch from the data stream, which should leave stationary Gaussian noise if there are no other glitches or signals in the data segment. We can use this expectation as a test to quantify the quality of the subtraction. Once we verify that we have faithfully modelled the glitches, we will investigate the distribution of parameters over the population of each type of glitch.

Our model is a matched filter for the glitch and with the parameter space of $f_0$ and $\gamma$ forms a manifold of templates. We can therefore define the optimal matched-filter SNR of a glitch as well as the match between any two glitch templates -- or the match with other types of waveforms like compact binary coalescence templates. We first define a scalar product
\begin{equation}
\langle a | b \rangle = 4~\mathrm{Re} \mathlarger{\int}_{f_{min}}^{f_{max}} ~ \frac{a^*(f)~b(f)}{S_n(f)} df ~,
\end{equation}
with $f_{min}$ and $f_{max}$ the range of sensitive frequencies of the detector (roughly $10$ Hz to a few kHz). Though we write this as an integral, it is straightforward to discretize.
The SNR is 
\begin{equation}
\rho(t) = \frac{1}{\sigma}~\langle h(f) e^{-2 \pi i f t}~|~d(f) \rangle ~,
\end{equation}
where $\sigma^2 = \langle h|h \rangle$ is a normalization factor. We have written the time-shift operator $e^{-2 \pi i f t}$ explicitly here but it can also be considered part of the template.
The match of two waveforms $a$ and $b$ is
\begin{equation}
\mathcal{M}(a, b) = \max_{t, \phi} ~ \frac{\langle a | b \rangle}{\sqrt{\langle a | a \rangle ~ \langle b | b \rangle}}
\end{equation}
where the maximum is over the relative time and phase of the two templates. Because our glitch template has no phase evolution, the maximum over time and phase is unnecessary, making the match between any two of our glitch templates very cheap to compute if needed.

\begin{figure}[!htbp]
\includegraphics[width=1.0\columnwidth]{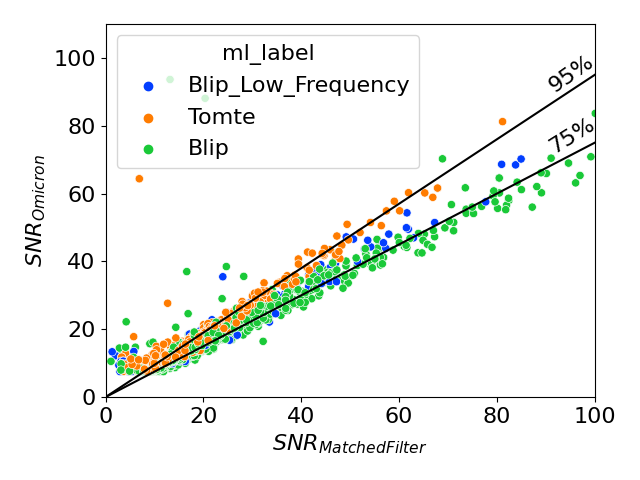}
\caption{The SNR reported by Omicron compared to the matched filter SNR from our model template. As expected, the value is lower; the lines mark the fraction of recovered SNR.}
\label{fig:SNRMatchGSpy}
\end{figure}

We define the SNR of a glitch in real data as the maximum of the SNR over our model parameters. We calculate this value using the maximum-likelihood results. Fig.~\ref{fig:SNRMatchGSpy} compares this matched-filter SNR to the SNR quantity reported by GravitySpy (and computed by Omicron). No single Q tile matches the glitch as well as our model template, so the Omicron SNR is always lower. Different types of glitches retain different fractions of the total SNR, as indicated by the lines.

When quantifying the effectiveness of our model at removing each glitch we do not use the SNR, as it assumes that the glitch matches the template. Instead, we use an excess power statistic which is agnostic to the waveform of the glitch. 
The power $P$ is computed as a sum of squares of the whitened times series of the central 16th of a second, divided by the expectation value. The whitened noise, with no glitch or signal present, should be a sample of unit Gaussians; hence the power follows a scaled chi-squared distribution with $N$ degrees of freedom and scaled by $1/N$, where $N$ is the number of time samples used. With only noise present (no glitch or signal), $P$ has mean $1$ and variance $\sqrt{\frac{2}{N}}$. In our case, $N = 8192/16 = 512$.
We find that the power and matched filter SNR are well-fit by
\begin{equation}
\rho = \sqrt{N (P - 1)}
\label{eq:SNRvsPower}
\end{equation}
for the majority of glitches.

\subsection{Residuals}

\begin{figure*}[!htbp]

\centering
\begin{tabular}{cc}
\includegraphics[width=1.0\columnwidth]{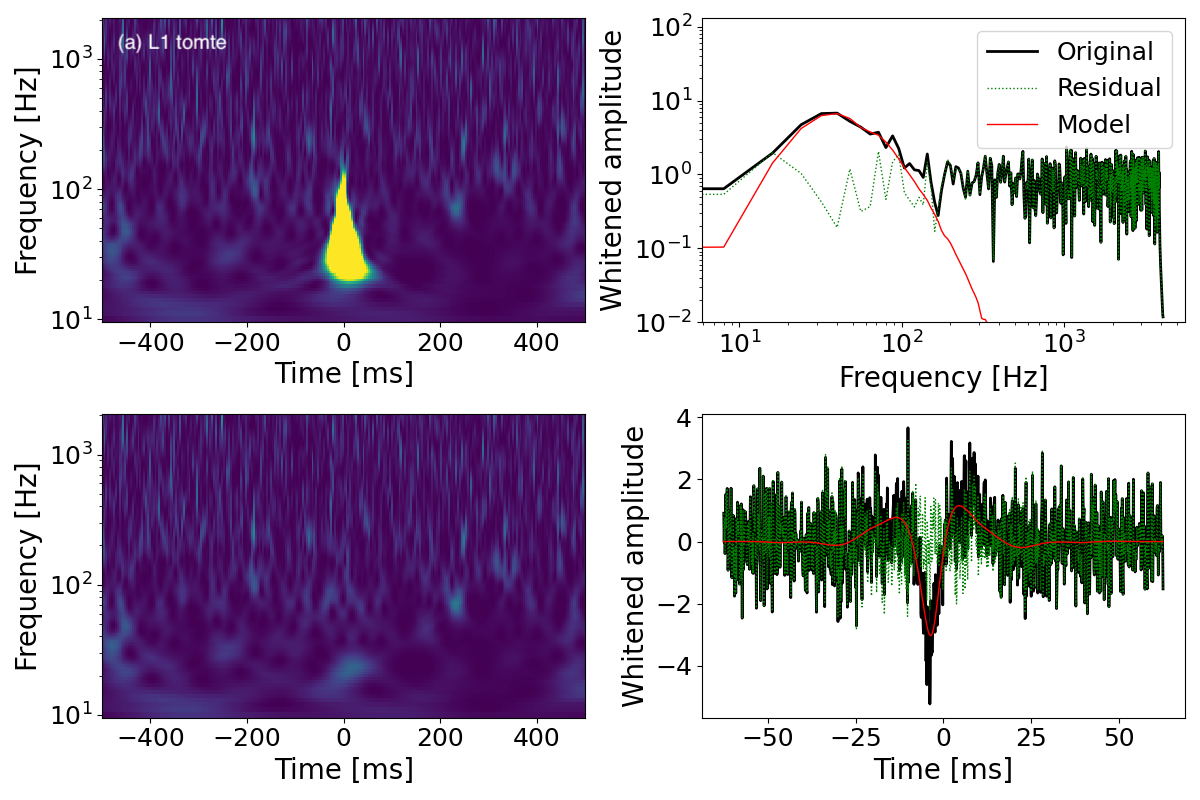}&
\includegraphics[width=1.0\columnwidth]{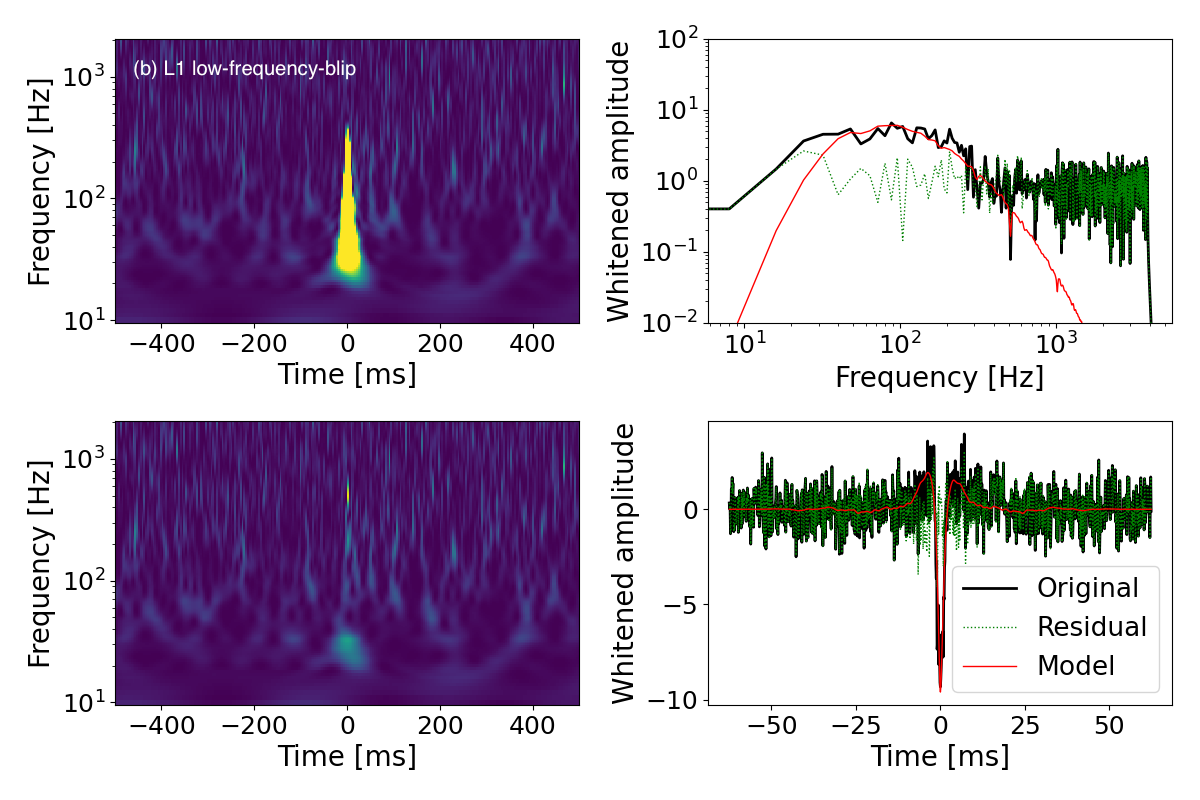} \\ [2\tabcolsep]
\includegraphics[width=1.0\columnwidth]{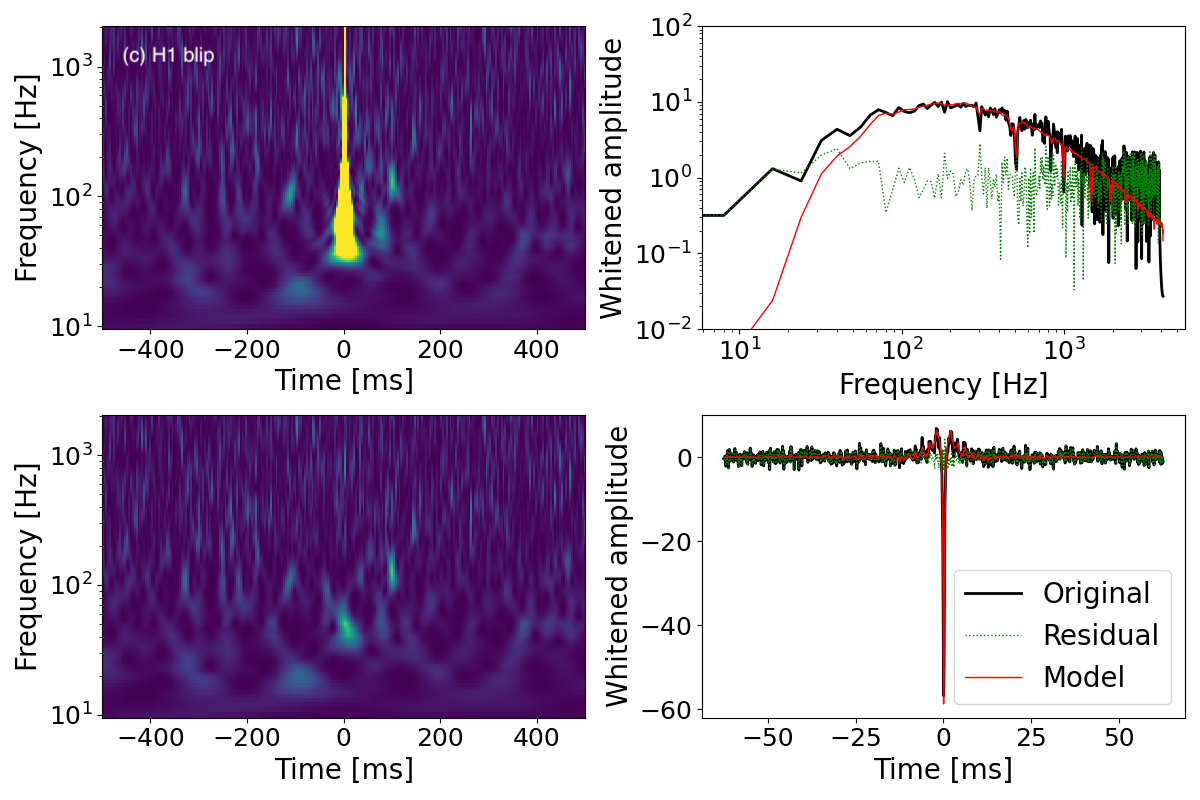}&
\includegraphics[width=1.0\columnwidth]{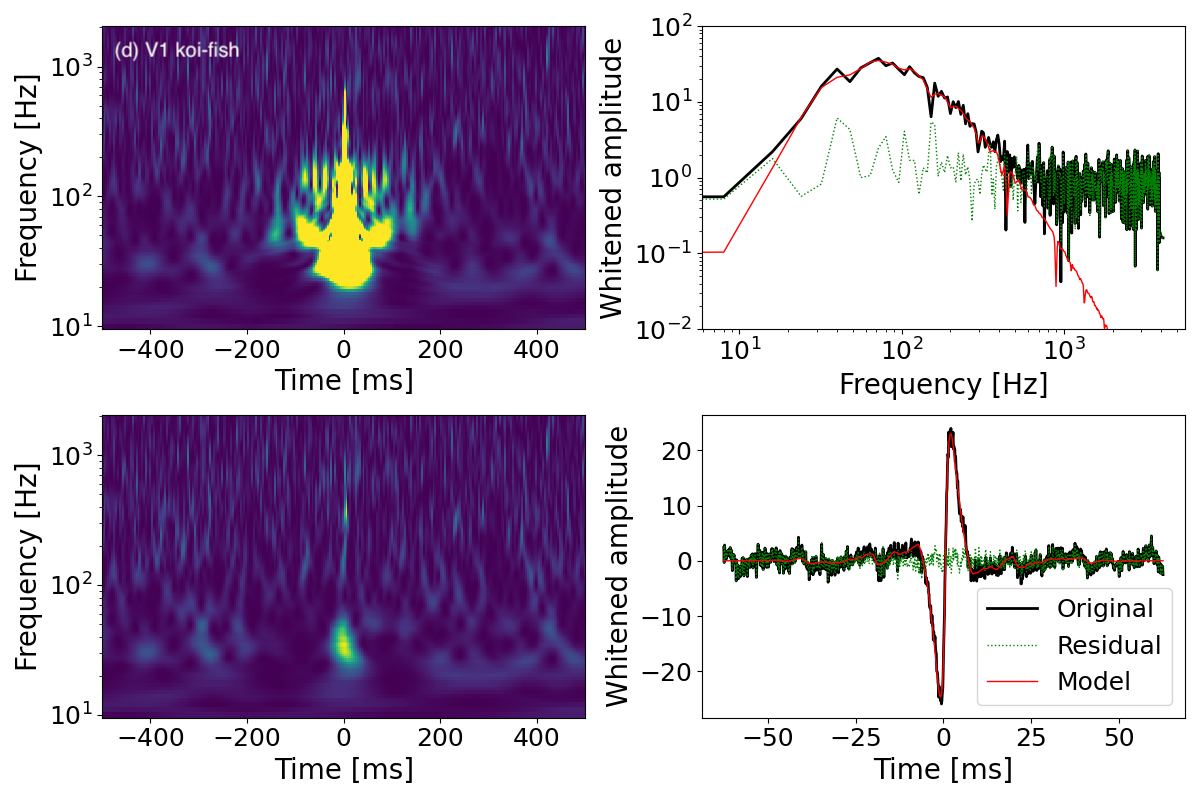}
\end{tabular}
\caption{(a) Tomte removal in L1, (b) Low-frequency blip removal in L1, (c) blip removal in H1 and (d) koi-fish glitch removal in V1. In each subfigure, the left panel shows the spectrogram before (top panel, left) and after subtraction (bottom panel, left), while the right panel displays the strain in frequency (top panel, right) and time domain (lower panel, right). The fit (red) is subtracted from the data (black) leaving behind the residual (green). Each case is an example of a successful subtraction with a residual  $\sim 1$, i.e., noise level.    }
\label{fig:SpectrogramRemoval}
\end{figure*}

We visualize the effectiveness of the subtraction in several different ways. We can compare the Q scan spectrograms before and after the subtraction. Fig. \ref{fig:SpectrogramRemoval} shows the effect of the subtraction for sample tomte, blip, low-frequency blip and koi fish glitches. For each glitch, the Q scan is shown before and after the subtraction on the left, with a fixed colour scale for the excess power statistic. The removal is also shown for whitened data in both the frequency and time domains. The expected spectrum in just noise should be a straight line at a value of $1$. The examples in Fig. \ref{fig:SpectrogramRemoval} are chosen to be representative glitches of the class that also have good removals.

\begin{figure}[!htbp]
\includegraphics[width=1.0\columnwidth]{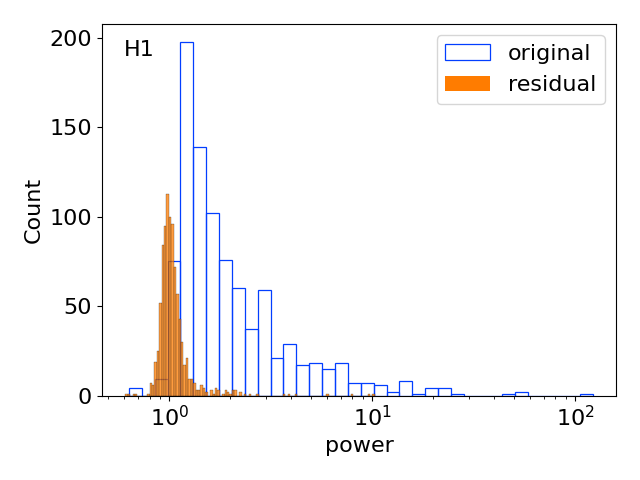}
\includegraphics[width=1.0\columnwidth]{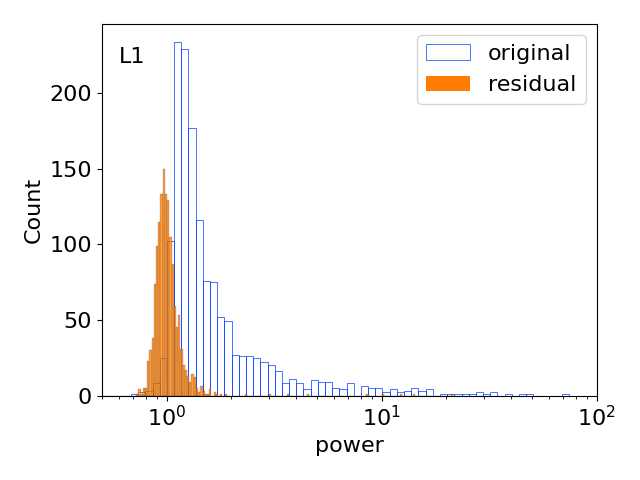}
\includegraphics[width=1.0\columnwidth]{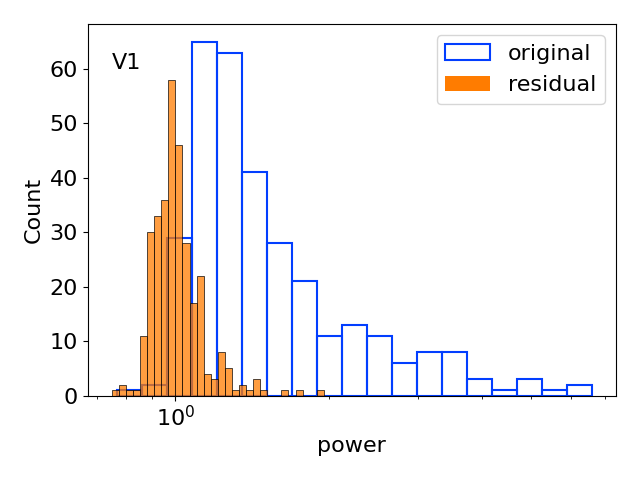}
\caption{The original and residual power after glitch removal is shown for L1, H1 and V1.  The residual centers at $1$ as expected for noise. The koi fish glitches will be displayed separately.}
\label{fig:ResidualPower}
\end{figure}

\begin{figure}[!htbp]
\includegraphics[width=1.0\columnwidth]{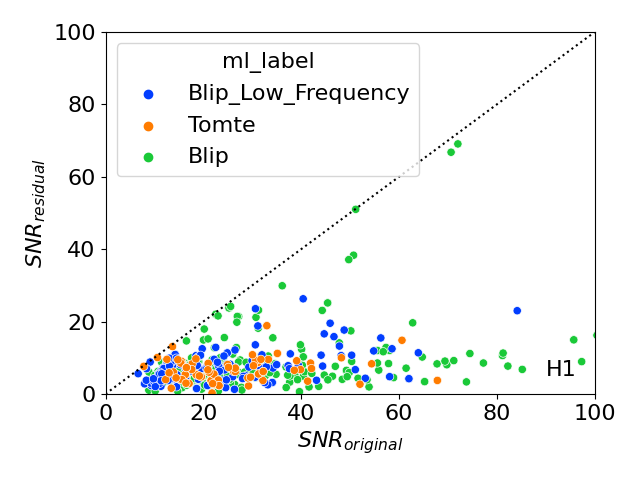}
\includegraphics[width=1.0\columnwidth]{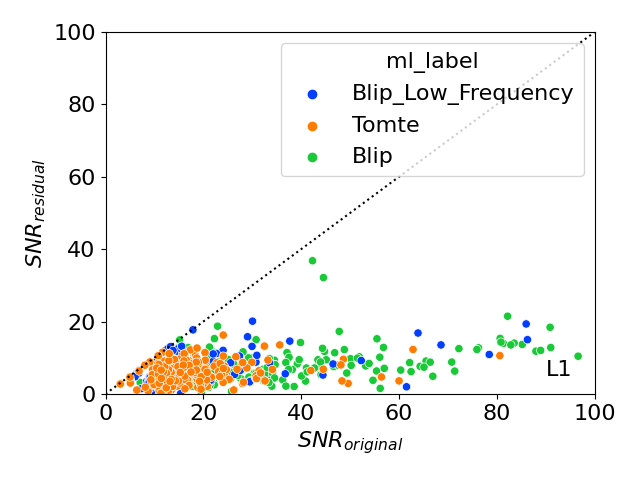 }
\includegraphics[width=1.0\columnwidth]{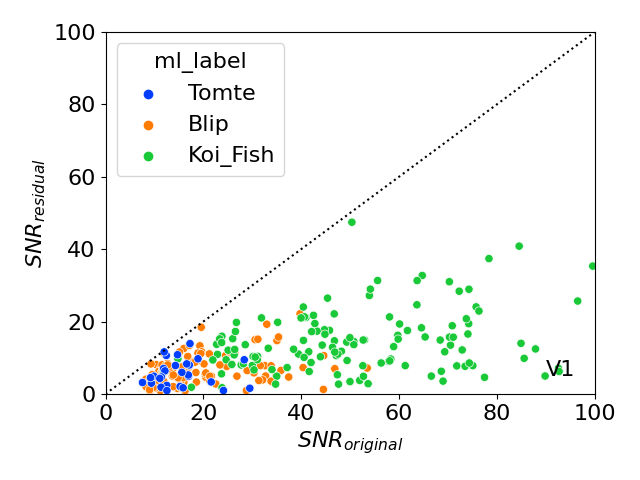 }
\caption{The residual SNR obtained after glitch removal as a function of the original SNR for H1, L1 and V1. The solid line indicates no reduction.}
\label{fig:SNROriginalResidual}
\end{figure}

\begin{figure}[!htbp]
\includegraphics[width=1.0\columnwidth]{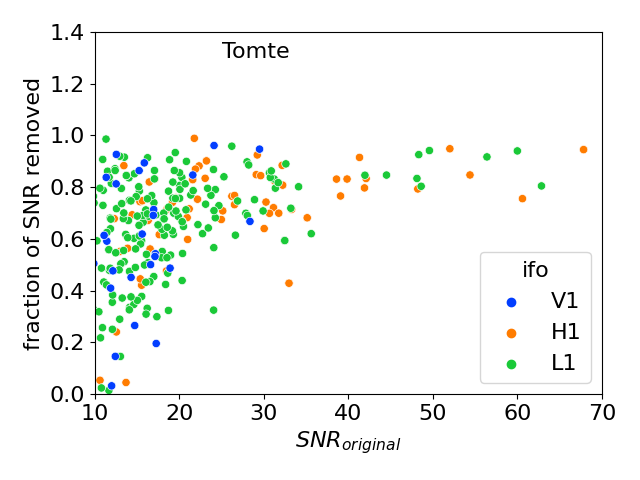}
\includegraphics[width=1.0\columnwidth]{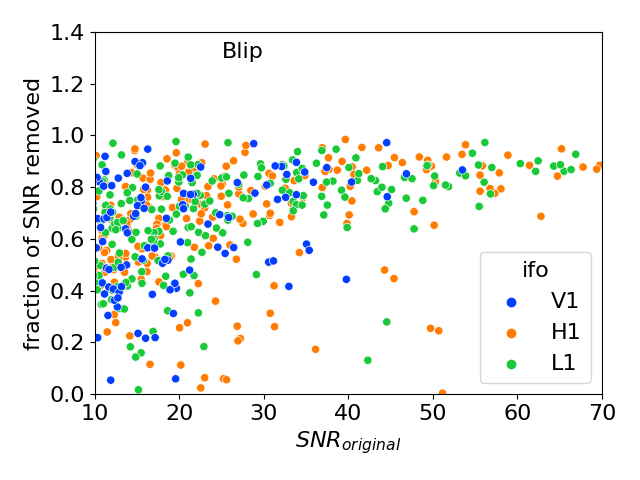}
\includegraphics[width=1.0\columnwidth]{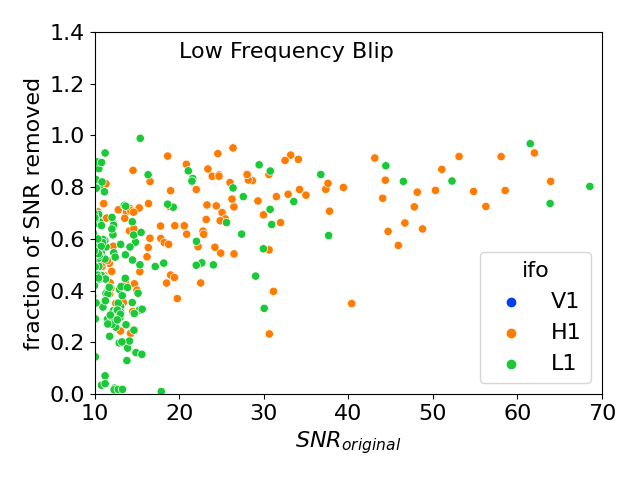}
\caption{The fractional SNR removed for different types of glitches. Proximity to 1 suggests complete removal. Most glitches have been removed with a few outliers for each of the glitch types.}
\label{fig:SNRFraction}
\end{figure}

To test the reliability of the subtraction on the entire dataset, we compute the residual power after each glitch subtraction. An ideal removal would result in $P$ consistent with the expected scaled chi-squared distribution. Fig. \ref{fig:ResidualPower} shows the residual power compared to the power before removal. We will consider the koi fish separately due to their much larger amplitude distribution.
Fig. \ref{fig:SNROriginalResidual} displays the residual SNR 
calculated using Eq. \ref{eq:SNRvsPower} versus the original matched-filter SNR. The distribution appears to be independent of the original SNR except for a small proportion of outliers. Fig. \ref{fig:SNRFraction} shows the fraction of the original SNR removed. This tends towards $1$ as the original SNR increases, while at a lower SNR the distribution is wider. This can be due to both of the SNRs being relatively more uncertain at low values, and the removal working less well because noise has a greater effect on the inference of the parameters.

After removal, the residual was greater than $2$ for 16 out of 442 total blips in H1, 9 out of 500 blips in L1, 0 out of 264 blips in V1, and for none of the tomtes (of 500 in L1, 135 in H1, and 53 in V1). Several of these ineffective removals were due to GravitySpy misidentifications: some had multiple glitches rather than an isolated glitch, and there were at least three RF whistle glitches misclassified as blips.

\begin{figure}[!htbp]
\includegraphics[width=1.0
\columnwidth]{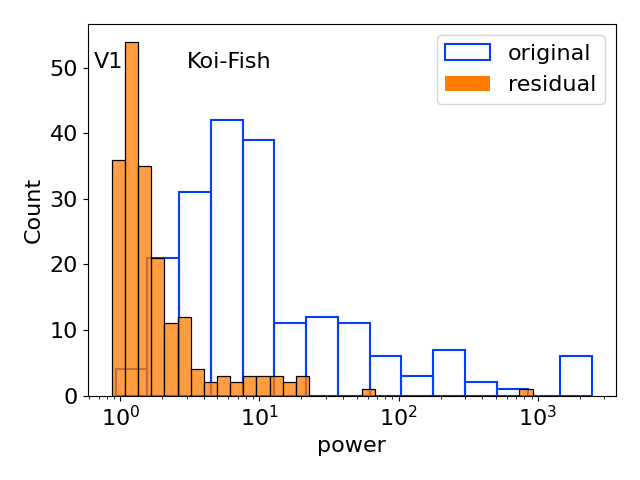}
\includegraphics[width=1.0\columnwidth]{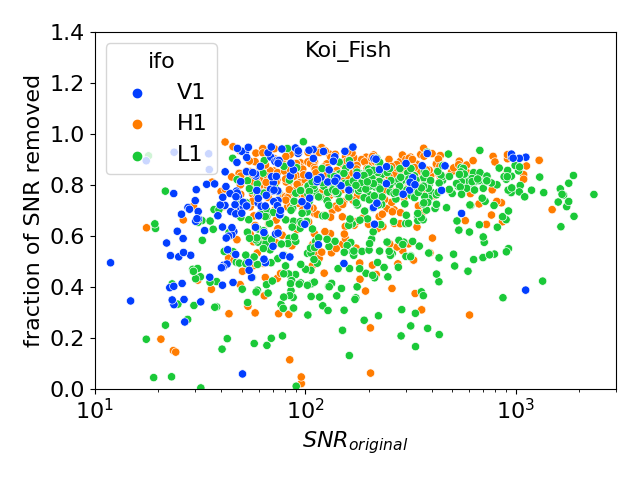}
\caption{(a) The original and residual power after glitch removal is shown for koi fish glitches in V1. (b) Fraction of SNR removed in L1, H1 and V1 for koi fish glitches}
\label{fig:ResidualPowerKoi}
\end{figure}

The removal of koi fish glitches was far less effective than the other types. Fig. \ref{fig:ResidualPowerKoi}(a) shows the residual power for V1 koi fish. There is a long tail, with 51 out of 196 glitches above a residual power of $2$.
Fig. \ref{fig:ResidualPowerKoi}(b)) shows the fractional removal for all three detectors. The removal is worse for the LIGO detectors, with half of the koi fish above a residual power of $2$.
Some of these poor removals are caused by the presence of multiple glitches when they should be isolated. But the removal is imperfect in cases of true koi fish glitches as well. The koi fish glitch has a more complicated structure that is not as well captured by our model as for the other types.

\subsection{Inferred Parameters}

\begin{figure}[!htbp]
\includegraphics[width=1.0
\columnwidth]{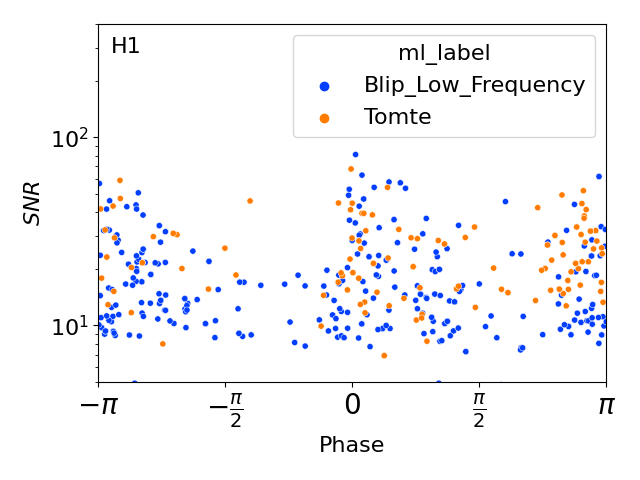}
\includegraphics[width=1.0
\columnwidth]{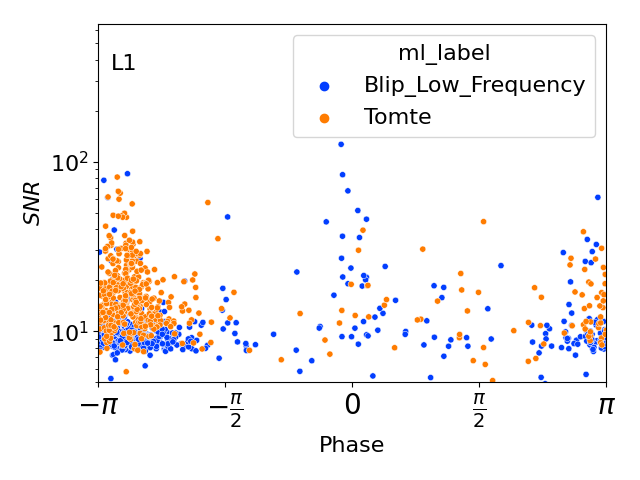}

\caption{The SNR vs phase for a low-frequency blip and tomte glitches in H1 and L1. The glitches cluster at $\pm \pi$ and $0$ (symmetric). In L1, the glitches cluster predominantly at  $-\pi$.} 
\label{fig:SNRPhaseBlipLfTomte}
\end{figure}

\begin{figure}[!htbp]
\includegraphics[width=1.0
\columnwidth]{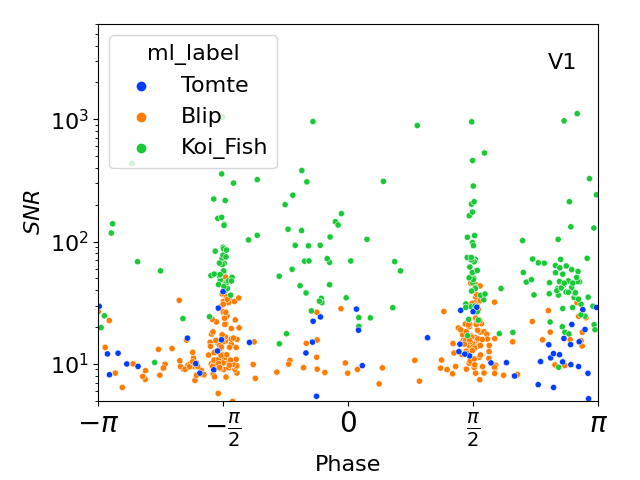}
\caption{The SNR vs phase for all glitches in V1. They cluster at $\pm \pi/2$, and are thus anti-symmetric in time.} 
\label{fig:SNRPhaseV1}
\end{figure}

\begin{figure}[!htbp]
\includegraphics[width=1.0
\columnwidth]{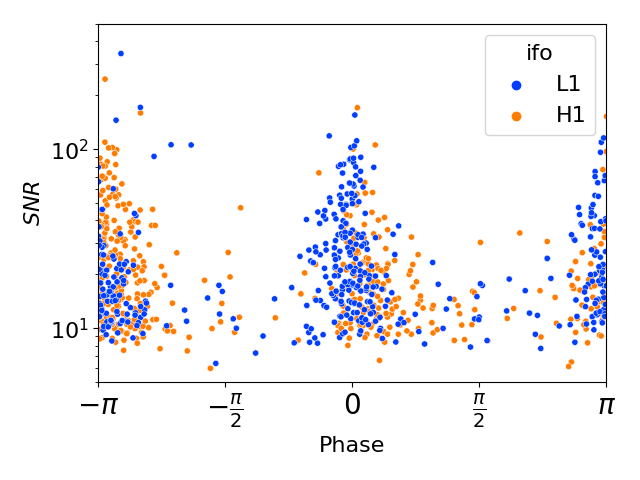}
\caption{The SNR vs phase is shown for blip glitches in H1 and L1. Unlike the tomtes, they appear to cluster evenly in phase at $0$ and $\pm \pi$.} 
\label{fig:SNRPhaseBlip}
\end{figure}

From our fits, we have access to the distribution of the parameters of all of the glitches for which our model worked well. Here we consider mainly the distribution of SNR and phase. The low-frequency blips and tomtes have very similar parameters, except the tomtes have slightly higher SNR. In Fig. \ref{fig:SNRPhaseBlipLfTomte}, the phase and SNR of low-frequency blips and tomtes are compared, for each detector. L1 has far more of these two types than the other detectors. The tomtes all have a phase close to $-\pi$ (i.e. a negative symmetric peak in time). Low-frequency blips at low SNR mimic this, but there is another population at a phase of $0$ with higher SNR. These are likely to be a different population with a different cause. In H1, there are clusters around both $0$ and $\pi$, but the clusters have more variance. In both cases, the phase varies around these values less at high SNR, but still enough to indicate intrinsic variance rather than the effect of noise.

Fig. \ref{fig:SNRPhaseV1} displays the results for all glitches in V1. Blips and koi fish cluster around a phase of $\pm \pi$, which are glitches anti-symmetric in time. There is another weaker cluster possible around $\pi$. This likely indicates a different mechanism from the one causing the glitches in LIGO, though the similarity of blips and koi fish indicates that they may have a common cause, and only the amplitude is different.

Fig. \ref{fig:SNRPhaseBlip} shows that the blips in H1 and L1 have similar characteristics, with clusters at phases of both $0$ and $\pi$ which have equally strong SNRs. Most coherent signals are expected to appear with opposite phases in H1 and L1 since they are nearly co-aligned but mirrored, so there is a danger of a blip with one phase accidentally appearing coincident with a blip of the opposite phase in the other detector.

\section{Match with Compact Binary Templates}

\begin{figure}[!htbp]
\includegraphics[width=1.0
\columnwidth]{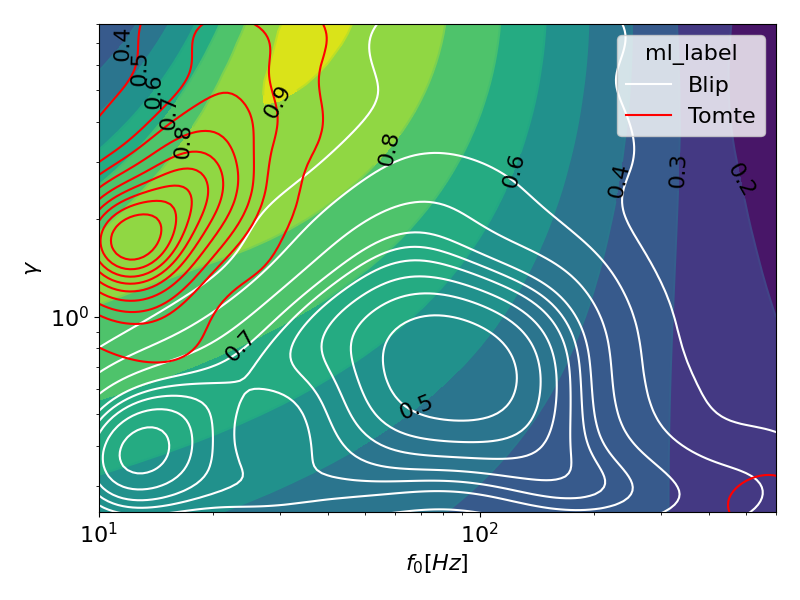}
\caption{The distribution of tomte (red) and blips (white) on the best match computed when maximizing over total mass, spin and mass ratio. The contours show that the blip distribution peaks in the low match region (${\cal M} \approx$0.4-0.5), while the tomte peak at the high maximum match (${\cal M} \approx0.8$) and are thus more likely to be mistaken for high mass binary black holes.} 
\label{fig:MatchBlipTomte2}
\end{figure}

Our quasi-physical model allows direct calculation of the impact that these glitches have on a compact binary coalescence (CBC) search. We compute the match between CBC templates and a glitch template. For a glitch of SNR $\rho_g$ in the data, the CBC template will have a peak of SNR $\rho_b = \mathcal{M}_{b,g} \rho_g$, where $\mathcal{M}_{b,g}$ is the match. In addition, the higher the match, the more difficult it is to distinguish whether a glitch or a CBC signal is present, even with an optimal method.

Fig. \ref{fig:MatchBlipTomte2} shows the distribution of observed blip and tomte glitches in LIGO over the model parameters $f_0$ and $\gamma$. Under these contours, the maximum match over the CBC template bank with a glitch template is displayed. We use the standard IMRPhenomXAS \cite{imrphenom} template and a model O3 noise curve. The center of the blip distribution is in a region where the match is only $0.5$, so these glitches should be difficult to mistake for real CBC signals. However, the tomtes range over a region where the match is between $0.8$ and $0.9$, making them very similar to CBC templates.

\begin{figure}[!htbp]
\includegraphics[width=1.0
\columnwidth]{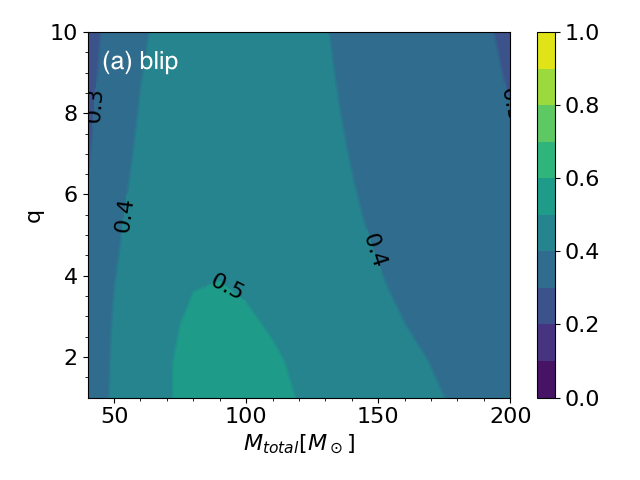}
\includegraphics[width=1.0
\columnwidth]{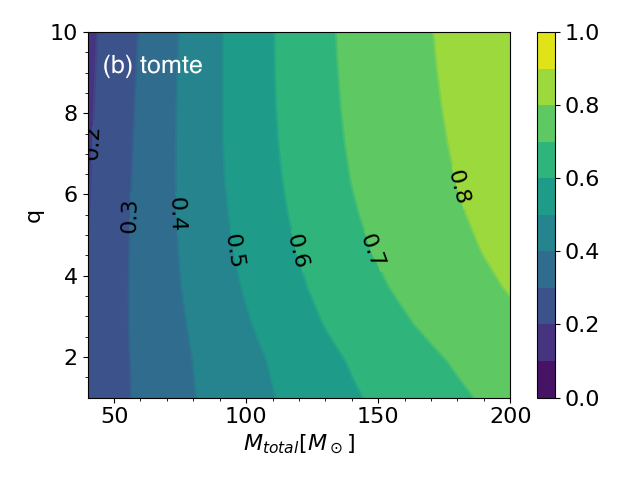}
\includegraphics[width=1.0
\columnwidth]{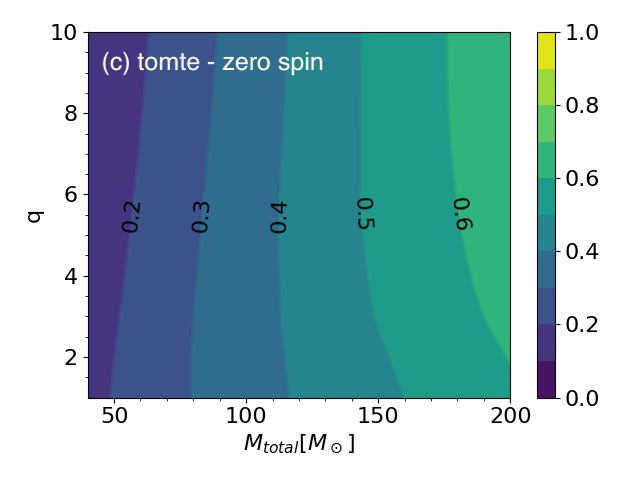}
\caption{Matching (a) blip and (b) tomte glitches to standard IMRPhenomXAS templates, we see that the match with the Blip glitches is generally low and thus mistaking them for high mass black holes can be avoided, while tomte glitches more easily be mistaken for high mass black holes. The maximum match favours anti-align spin. In (c) the average tomte is matched for zero spin. Fixing the spin lowers the match significantly, i.e., by about 0.2. Furthermore, the match becomes independent of q even for high $M_{\rm total}$.} 
\label{fig:MatchBlipTomte1}
\end{figure}

Fig. \ref{fig:MatchBlipTomte1} shows the match as a function of CBC parameters for a typical blip and tomte glitches. The overlap of blips with CBC templates peaks at $M_{t} = M_1 + M_2$ of $90$. The match peaks at equal masses, though not strongly. The overlap of tomtes with CBC templates peaks at $M_{t} = M_1 + M_2$ of $160 - 200$, depending on the parameters of the glitch. Here, asymmetric mass ratios are slightly preferred. In both cases, the highest matches occur with a negative aligned spin. However, Fig. \ref{fig:MatchBlipTomte1}c shows that tomtes still produce high matches even when the CBC templates are restricted to zero spin. These results were calculated using the Model O3 PSD for the LIGO detectors.

\section{Discussion}
\subsection{Sketch of practical use}

The two primary uses of our quasi-physical glitch model are to reliably detect the presence of a glitch and to precisely subtract the glitch from the data. In this work, we have used Omicron and GravitySpy to locate and do the initial classification of glitches. But that could be replaced by a matched-filter search using a template bank of model glitches covering the range of observed parameters. The matched filter gives a more accurate SNR and can predict the effect on a CBC search. Also, comparing the SNRs for each alternative can test whether a candidate event is a CBC or a glitch. This is more optimal than methods that do not have a matched-filter model for the glitch.

Whenever a glitch is detected, it is simple to remove. The maximum likelihood estimate can be calculated in just seconds. The residual data virtually eliminates the presence of the glitch in the case of blips, low-frequency blips, and tomtes. Koi fish will need further work and likely a more complicated model to be subtracted effectively, but they can at least be easily identified.

Our model can also be used in Bayesian parameter estimation of CBC signals that overlap glitches. The glitch model can be added to the likelihood and the glitch parameter space adjoined to the prior to allow simultaneous inference and disentangling of the two transients. Because our model is effective with a low-dimensional parameter space, it should not excessively affect the runtime of the parameter estimation.

\subsection{Possible Glitch Mechanisms}

For the most part, the physical cause of these short glitches is unknown. There have been extensive attempts to find correlations with auxiliary witness sensors, but for the most part, the glitches occur with no trace in any of these systems. Some blip glitches in previous observing runs were caused by delays in inter-computer communication in the control system \cite{cabero2019blip}. Others are correlated with low humidity, possibly caused by static discharge in the electronics \cite{humidity1, humidity2}. Before and during O4, there are tomte or low-frequency blip glitches caused by large electrical power draws \cite{electricalO4}.

Our quasi-physical model of glitches models the spectrum and also the time series of the detector strain. A hypothetical mechanism for a glitch can hence be compared to the physical motion of one of the test mass mirrors that would be implied. For instance, the fact that LIGO blips are equally likely to have either a positive or negative sign could indicate that they arise from the motion of either of the end test masses.  

For glitches produced in another part of the sensing chain (not as physical motion of the mirrors), the implied disturbance can be reconstructed by modelling the transfer function to strain. For instance, a disturbance in the light reaching the readout photodiode will affect the measured strain through the optical sensing function. The phases seen in Virgo could indicate a derivative coupling as this produces a phase shift of $\pi/2$.

The amplitude and SNR distributions are also useful clues to the origin of these glitches. It is difficult to identify glitches at low SNR, and the GravitySpy results have a cutoff at SNR $7.5$. We see that the SNR distributions are peaked toward low values (except for koi fish), and are truncated by the threshold. This suggests that a large number of these glitches are not identified because of the cutoff. The matched filter model presented here can be used to detect and analyze these numerous glitches at low SNR.

\section{Conclusions}
LIGO has recently begun its fourth observing run and is producing the most sensitive data of the gravitational wave sky recorded to date. A limiting factor in identifying gravitational wave signals continues to be transient non-Gaussian noise that mimics signals, which are classified as glitches. In this paper, we have modelled and removed short transient glitches which occur often in both the LIGO and Virgo detectors.

Although we do not have a detailed physical model of the glitch process, we are able to create a parameterised frequency-domain model based on their properties. The glitches are very short with less than two cycles of oscillation with power concentrated around their central time. This is well-fit by a normal in log-frequency with four parameters: the peak frequency, bandwidth, amplitude, and phase. Our model is tested on blip, tomte, low-frequency blip, and koi fish glitches as classified by GravitySpy. We demonstrate the accuracy of the model and the effectiveness of glitch subtraction by testing on one month of O3 data. More than 90\% of blips, tomtes and low-frequency blips were successfully removed. The removal of koi fish glitches is less effective because they have a more complicated structure and also a very high amplitude.

We match the phase space of tomte and blip glitches to binary black hole merger templates and find that most blip glitches have a high mismatch of around $0.5$ with high mass black holes modelled by the IMRPhenomXAS template. On the other hand, tomte glitches can be more easily mistaken for high-mass black holes with a typical mismatch of $0.2$. The primary difference between the black hole waveforms and the glitch waveform is the lack of phase of the evolution of the latter.

We provide a sketch of practical uses of our glitch model in black hole searches and parameter estimation and for investigation of the causes of these glitches, which can be used to mitigate or eliminate them.

\section*{Acknowledgements}
\label{Acknowledgements}
The authors are grateful to the members of the Gravitational Waves research group from ICC UB for their continual advice and support. We are particularly grateful to our internal reviewer, Francesco DiRenzo. We thank Mark Gieles, Tomas Andrade, Juan Trenado, and Nadejda Blagorodnova Mujortova from the ICC for useful discussions. We also thank ICC secretaries, Esther Pallarés Guimerà and Anna Argudo who go well beyond their job description to keep everything running smoothly. 

We acknowledge support from the Spanish Ministry of Science and Innovation through grant PID2021-125485NB-C22 and  CEX2019-000918-M funded by MCIN/AEI/10.13039/501100011033. RB acknowledges support from the AGAUR through grant SGR-2021-01069. AL and RM are supported by STFC grants ST/T000550/1 and ST/V005715/1.  This work has used the Python packages PyCBC \url{https://doi.org/10.5281/zenodo.7547919} and GWpy \url{https://doi.org/10.5281/zenodo.7997251}. The authors are grateful for computational resources provided by the LIGO Laboratory and Cardiff University and supported by National Science Foundation Grants PHY-0757058 and PHY-0823459 and STFC grant ST/I006285/1. Supporting research data are available on reasonable request from AL. For the purpose of open access, the author(s) has applied a Creative Commons Attribution (CC BY) licence to any Author Accepted Manuscript version arising.


\bibliographystyle{apsrev4-1}
\bibliography{References.bib}

\end{document}